\documentclass[aps,prc,amsmath,amssymb,twocolumn,superscriptaddress
]{revtex4-2}

\usepackage{graphicx}

\usepackage{braket}

\begin{document}


\title{Finite-amplitude method for collective inertia in spontaneous fission}


\author{Kouhei Washiyama}
\email[]{washiyama@phys.kyushu-u.ac.jp}
\affiliation{Research Center for Superheavy Elements, Kyushu University,
  Fukuoka 819-0395, Japan}

\author{Nobuo Hinohara}
\affiliation{Center for Computational Sciences, University of Tsukuba,
  Tsukuba 305-8577, Japan}
\affiliation{Faculty of Pure and Applied Sciences, University of Tsukuba,
  Tsukuba 305-8571, Japan}

\author{Takashi Nakatsukasa}
\affiliation{Center for Computational Sciences, University of Tsukuba,
  Tsukuba 305-8577, Japan}
\affiliation{Faculty of Pure and Applied Sciences, University of Tsukuba,
  Tsukuba 305-8571, Japan}



\date{\today}

\begin{abstract}
\begin{description}
\item[Background]
  Microscopic description of spontaneous fission
  is one of the most challenging subjects in nuclear physics.
  It is necessary to evaluate the collective potential and
  the
  collective inertia
  along a fission path for a 
  description
  of quantum tunneling
  in spontaneous or low-energy fission.
  In 
    past studies of the fission dynamics 
  based on nuclear energy density functional (EDF) theory,
  the collective inertia has been evaluated with the cranking approximation,
  which neglects dynamical residual effects.
%
\item[Purpose]
  The purpose is
  to provide a reliable and efficient 
  method 
	to include dynamical residual effects
  in the collective inertia for fission dynamics.
\item[Methods]
  We use the local quasiparticle random-phase approximation (LQRPA)
  to evaluate the collective inertia along a fission path
  obtained by the constrained Hartree-Fock-Bogoliubov method
  with the Skyrme EDF.
  The finite-amplitude method (FAM) with a contour integration technique
    enables us to efficiently compute
  the collective inertia in a large model space.
\item[Results] We evaluate the FAM-QRPA collective inertia
  along a symmetric fission path in $^{240}$Pu and $^{256}$Fm.
  The FAM-QRPA inertia 
		is significantly larger than the one of the cranking approximation, and
		shows pronounced peaks around the ground state and
  the fission isomer.
  This is
  due to dynamical residual effects.
%
\item[Conclusions]
  To describe the spontaneous or low-energy fission, 
  we provide a reliable and efficient 
    method to construct the collective inertia
		with dynamical residual effects that have been neglected in 
		most of EDF-based works in the past.
	        We show the importance of 
                dynamical residual effects to the collective inertia.
    This work will be a starting point for a systematic study
    of fission dynamics in heavy and superheavy nuclei
    to 
    microscopically describe the nuclear large-amplitude collective motions.
 
\end{description}
\end{abstract}


\maketitle

\section{Introduction}


%
Nuclear fission~\cite{schunck16, andreyev17}
plays an important role in various phenomena
such as synthesis of superheavy elements~\cite{oganessian15,hofmann15}
and $r$-process nucleosynthesis in the universe~\cite{mumpower16,giuliani18a}. 
The 
  fission governs the existence and decay property of superheavy nuclei.
  Various 
types of fission, such as neutron-induced fission, spontaneous fission, and
$\beta$-delayed fission that involve neutron-rich heavy and superheavy nuclei,
are important in expected environment in the $r$-process.
%


%
Theoretically, a microscopic description of
large-amplitude collective motions, such as nuclear fission and fusion,
is one of the challenging subjects in quantum many-body physics. 
Of particular interest is a quantum many-body tunneling
for spontaneous or low-energy fission and subbarrier fusion of complex nuclei.

The most promising candidate for the microscopic approach is the
selfconsistent nuclear energy density functional (EDF) theory~\cite{bender03,nakatsukasa16}.
It is well established that 
the nuclear EDF gives a good description of ground-state properties
of nuclei in the whole nuclear chart.
The EDF approaches to the spontaneous fission
have used a semiclassical description with the WKB approximation for quantum tunneling.
These 
studies show how the result depends on the choice of 
relevant collective variables, 
the collective potential energy, and the collective inertia 
entering the action integral in the WKB approximation. 
%
The potential energy has been calculated by the Hartree-Fock-Bogoliubov method
with constraints on the collective variables.
%
On the other hand,
the collective inertia used
in the previous EDF-based works~\cite{baran11, sadhukhan13, sadhukhan16, sadhukhan17,giuliani18b}
is insufficient in the following respect: 
First of all,
the previous works employ the so-called cranking approximation 
to the adiabatic time-dependent Hartree-Fock-Bogoliubov (ATDHFB) method \cite{baranger78}
to evaluate the collective inertia.
The cranking approximation neglects dynamical residual effects, especially, the 
time-odd terms of the EDF.
This gives a big drawback, 
namely, the violation of the (local) Galilean symmetry,
which leads to the incorrect total mass for the translational motion
\cite{ring-schuck,nakatsukasa16}.
The rotational moments of inertia of deformed nuclei are not properly given in the
cranking approximation~\cite{thouless62}.
In addition, the
collective inertia 
  of the cranking approximation
  significantly deviates from 
  the one including
the dynamical residual effects~\cite{dobaczewski81}.
%
%
%
Irrespective of such drawbacks, 
  because of its simplicity, 
the cranking approximation has been widely used in evaluating collective inertia
not only for fission dynamics
but also for the collective Hamiltonian method~\cite{libert99,prochniak04,niksic09,delaroche10,li10}.
It should be noted that the enhancement factors of $1.2$--$1.4$ are
often adopted to correct the missing residual effect.
%


An alternative
method to ATDHFB for describing large-amplitude collective motions
has been developed,
called the adiabatic selfconsistent collective coordinate (ASCC) method~\cite{matsuo00}.
Based on
the ASCC method, an efficient 
  and feasible method to construct
the collective Hamiltonian
has been 
proposed~\cite{hinohara10}.
In this method,
  the collective path (subspace) is determined by the constrained HFB (CHFB) calculation,
while 
the collective inertia (the vibrational mass in the vibrational kinetic Hamiltonian)
is determined by local normal modes built on 
CHFB states.
Local normal modes are obtained by solving local quasiparticle random-phase approximation (LQRPA) equations.
Dynamical residual effects with time-odd components are
selfconsistently included in the collective inertia by the LQRPA.
This method, called CHFB + LQRPA, has been applied to constructing
five-dimensional quadrupole collective Hamiltonian for triaxial shapes
with the pairing-plus-quadrupole (P+Q) Hamiltonian~\cite{hinohara10, sato11, hinohara11, sato12},
a hybrid model of the P+Q Hamiltonian and covariant EDF~\cite{hinohara12},
and three-dimensional quadrupole collective Hamiltonian for axially symmetric shapes
with the Skyrme EDF~\cite{yoshida11}.
Those works have shown 
the importance of including dynamical residual effects with time-odd components
on the description of large-amplitude collective motion.


Despite 
many evidences for the importance of dynamical residual effects,
no practical application based on the ATDHFB method takes into account residual effects
in the calculation of the collective inertia.
%
The reason 
is that it requires huge numerical computations
to apply directly the ATDHFB (or LQRPA) formulas to the collective inertia
in 
realistic
cases, due to the size of matrices treated 
with modern EDFs.
%

A method to efficiently solve the QRPA equations
based on linear response theory has been developed for the EDF,
called the finite-amplitude method (FAM)~\cite{nakatsukasa07}.
In the FAM, response functions to an external one-body field
can be obtained 
only with
one-body induced densities and fields with an iterative scheme.
%
%
The FAM requires matrices with the size of approximately $N\times N$,
while the QRPA equation requires 
those
with the size $\approx N^2 \times N^2$,
where $N$ represents the number of single-particle basis.
$N$ may exceeds $10^3$ 
when, for example,
20 major shells in the harmonic oscillator basis are used in solving deformed HFB equations. 
Thus, the computational cost of the FAM is significantly lower than 
that of QRPA.
As an alternative way 
of solving the QRPA equations, 
the FAM has been employed in various applications
\cite{inakura09, avogadro11, inakura11, stoitsov11, liang13, hinohara13, niksic13, mustonen14, pei14, hinohara15, kortelainen15, wen16, sun17, washiyama17}. 

The standard FAM formalism 
  has been extended to the one
in terms of the momentum--coordinate (PQ) representation
\cite{hinohara15b}
to treat zero-energy modes of 
QRPA solutions
known as spurious modes or Nambu--Goldstone (NG) modes~\cite{nambu60, goldstone61}
associated with spontaneous symmetry breaking of the mean field.
  A general method to calculate the inertia 
of the NG modes was also given~\cite{hinohara15b},
%
which was applied to pairing rotation~\cite{hinohara16,hinohara18}
and to spatial rotation of axially deformed 
\cite{kortelainen18}
and of triaxially deformed nuclei~\cite{washiyama18}.
It is important to note that 
this formulation can treat imaginary solutions of the QRPA
that may appear in the LQRPA 
at nonequilibrium HFB states
obtained by the CHFB calculations.
%

The main aim of this paper is
to propose a 
new method
to efficiently evaluate the collective inertia used in
the WKB approximation for spontaneous fission
and in the vibrational kinetic 
terms
in the collective Hamiltonian method.
%
The 
  method 
is based on
the CHFB + LQRPA with the Skyrme EDF.
We employ 
the PQ representation of the 
FAM
to efficiently solve the LQRPA equations.
We extend a contour integration approach of the FAM proposed in Ref.~\cite{hinohara13}
to
compute the collective inertia associated with
the most collective local normal mode in the LQRPA.

This paper is organized as follows.
Section II briefly summarizes the formulations of the collective inertia
in ATDHFB with the cranking approximation and in the LQRPA.
In Sec. III, we explain the formulation of the CHFB + LQRPA with the FAM
to calculate the collective inertia. 
In Sec. IV, we present the results of the collective inertia along mass-symmetric
fission path in ${}^{240}$Pu and ${}^{256}$Fm
and comparison between our results and those with the cranking approximation.
Conclusions are given in Sec. V.

\section{Collective inertia with EDF}

\subsection{ATDHFB}

We 
give a brief explanation for 
the collective inertia in ATDHFB.
For details, we refer to Refs. \cite{baranger78, dobaczewski81, baran11, giuliani18b}.

In ATDHFB, the collective motion of the system is 
assumed to be 
described by
a set of a few collective coordinates
$s_i$ 
and its conjugate momenta.
The collective coordinates are defined as 
$s_i=\langle \phi(s)|\hat{s}_i|\phi(s)\rangle$, 
where
$\hat{s}_i$ 
are generators driving the collective motion.
  In practice, the collective subspace $\ket{\phi(s)}$ is determined by
  the CHFB calculation with the constraint on $s_i$. 
Most applications take
$\hat{s}_i$ 
as one-body operators of multipole moments.
The ATDHFB relies on the assumption that
the collective velocities of the system denoted by the time derivative
of the collective coordinates,
  $\dot{s}_i$,
are slow enough to make adiabatic assumption valid.
The expression of the collective inertia tensor in ATDHFB is given by
\begin{equation}\label{eq:ATDHFB}
\mathcal{M}_{ij} = \frac{i}{2\dot{s}_i\dot{s}_j} \text{Tr}(F^{i*}Z^{j}-F^i Z^{j*}).
\end{equation}
The matrix $Z$ involves time-odd fields associated with
time-odd density matrices.
  A standard method to determine the matrix $Z$ requires
solution of the QRPA equation. 
The matrix $F$ is evaluated by
\begin{equation}\label{eq:fmatrix}
	\frac{F^{i}}{\dot{s}_i} = U^\dagger \frac{\partial \rho}{\partial s_i} V^*
	+ U^\dagger\frac{\partial \kappa}{\partial s_i} U^*
	- V^\dagger \frac{\partial \rho^*}{\partial s_i} U^*
	- V^\dagger \frac{\partial \kappa^*}{\partial s_i} V^*,
\end{equation}
where $U$ and $V$ are the Bogoliubov transformation matrices
and $\rho$ and $\kappa$ are the density matrix and pairing tensor, respectively.
This expression includes derivatives of density matrices with respect to
the collective coordinates.

\subsubsection{Cranking approximation} 

The cranking approximation 
neglects the time-odd fields
in the collective inertia \eqref{eq:ATDHFB}.
In this case, the matrix $Z^i$ can be simply written in terms of $F^i$ and
the quasiparticle energies.
%
The collective inertia tensor of the cranking approximation then reads
in the quasiparticle basis
\begin{equation}\label{eq:cranking}
	\mathcal{M}_{ij}^{\text{C}} = \frac{1}{2\dot{s}_i\dot{s}_j}
  \sum_{\mu\nu}\frac{(F_{\mu\nu}^{i*}F_{\mu\nu}^{j} + F_{\mu\nu}^i F_{\mu\nu}^{j*})}{E_{\mu}+E_{\nu}},
\end{equation}
where $E_{\mu}$ and $E_{\nu}$ are one-quasiparticle energies
defined with respect to the constrained Hamiltonian of the form of Eq.~(\ref{eq:CH})
and the CHFB state $\ket{\phi(s)}$ of Eq.~(\ref{eq:CHFB}).
%

\subsubsection{Perturbative cranking approximation to ATDHFB}

Further approximation leads to
the so-called perturbative cranking approximation,
where the derivatives of the densities
with respect to the collective coordinates in Eq.~\eqref{eq:fmatrix}
are not explicitly evaluated, but are obtained in a perturbative manner.
The expression of the collective inertia in the perturbative cranking approximation 
is given as
\begin{equation}\label{eq:perturb-crank}
	\mathcal{M}^{\text{PC}} = \frac{1}{2}
	[M^{(1)}]^{-1}M^{(3)} [M^{(1)}]^{-1},
\end{equation}
%
%
where the $n$-th energy-weighted moment $M^{(n)}$ is given as
\begin{equation}\label{eq:moment}
  M_{ij}^{(n)} = 
	\sum_{\mu <\nu} \frac{\langle \phi(s) |\hat{s}_i|\mu\nu \rangle
	\langle \mu\nu|\hat{s}_j^{\dagger}|\phi(s)\rangle}{(E_{\mu} + E_{\nu})^{n}},
\end{equation}
%
where 
$|\mu\nu\rangle$
are 
two-quasiparticle states
based on the CHFB 
state $\ket{\phi(s)}$. 

\subsection{CHFB + Local QRPA}\label{sec:LQRPA}

Another method to evaluate the collective inertia is
the CHFB + LQRPA \cite{hinohara10}
derived from the ASCC method~\cite{matsuo00}.
Suppose that a set of $N$ collective coordinates
$q_i$ ($i=1,\cdots,N$) 
selfconsistently determined with 
the ASCC method
can be mapped to a set of $N$ collective variables $s_m$ 
  ($m=1,\cdots,N$) 
through a one-to-one correspondence between them.
In addition, we assume that the ASCC collective subspace can
be approximated by the one obtained with the CHFB calculation with
a given set of constraining operators $\hat{s}_m$.
%
%
The CHFB equation is given by 
\begin{equation}
  \delta \langle \phi(s)| \hat{H}_{M} |\phi(s) \rangle = 0, 
	\label{eq:CHFB}
\end{equation}
with 
the constrained 
Hamiltonian
\begin{equation}
\hat{H}_{M} = \hat{H} - \sum_{\tau=n,p} \lambda_\tau \hat{N}_\tau - \sum_m \lambda_m \hat{s}_m ,
	\label{eq:CH}
\end{equation}
%
where $\lambda_{n,p}$ denote the Fermi energies for neutrons and protons
to constrain the average neutron and proton numbers
  $\bra{\phi(s)}\hat{N}_\tau\ket{\phi(s)}$ ($\tau=n,p$) 
and $\lambda_m$ are the Lagrange multipliers of constraining
$\langle \phi(s)| \hat{s}_{m} |\phi(s) \rangle=s_m$.
The energy minimization,
Eq.~(\ref{eq:CHFB}), 
leads to
the 
CHFB state $|\phi(s)\rangle$
and collective potential 
  $V(s)=\langle \phi(s)|\hat{H}|\phi(s)\rangle$. 
At
each CHFB state $|\phi(s)\rangle$,
the LQRPA equations,
\begin{align}
	\delta \langle \phi(s)| [\hat{H}_{M}, \hat{Q}_i(s)]
	- \frac{1}{i}\hat{P}_i(s) |\phi(s)\rangle &= 0, \\
	\delta \langle \phi(s)| [\hat{H}_{M}, \frac{1}{i}\hat{P}_i(s)] -
	\Omega_i^2(s)\hat{Q}_i(s) |\phi(s)\rangle &= 0,
\end{align}
are solved to determine
$\hat{Q}_i(s)$ 
and $\hat{P}_i(s)$ that are local generators
of collective momenta and coordinates, respectively, 
defined at $s$.
They should satisfy the weak canonicity condition,
Eq.~(\ref{eq:weak_canonicity_condition}).
%
Here, the inertial and the curvature tensors are diagonalized to define the local normal mode.
Furthermore, in order to fix the arbitrary scale of the collective coordinate $q_i$,
the inertia with respect to $q_i$ is set to be unity, ${\cal M}_{ij}=\delta_{ij}$.
$\Omega^2_i(s)$ 
denotes the squared eigenfrequency of the local normal mode. 
Note that the eigenfrequency of the LQRPA equations can be imaginary ($\Omega^2_i(s) <0$)
at 
nonequilibrium CHFB states.
A criterion for selecting relevant collective LQRPA modes from many LQRPA solutions
is given in Ref.~\cite{hinohara10}.



Once relevant LQRPA collective modes are selected, 
the collective inertia tensor is
given
as follows.
First, the collective kinetic energy is expressed as 
\begin{align}
	T= \frac{1}{2} \sum_i \dot{q}_i^2 = \frac{1}{2} \sum_{mn}{\cal M}_{mn}(s) \dot{s}_m\dot{s}_n,
\end{align}
where collective inertia tensor
  $\mathcal{M}_{mn}(s)$ 
is defined by
the transformation of the collective coordinates $q_i$
to the collective variables $s$, 
\begin{equation}\label{eq:vibrational-mass}
	\mathcal{M}_{mn}(s) \equiv
		\sum_{i,j} \frac{\partial q_i}{\partial s_m} {\cal M}_{ij} \frac{\partial q_j}{\partial s_n} =
	\sum_i \frac{\partial q_i}{\partial s_m} \frac{\partial q_i}{\partial s_n}.
\end{equation}
Second, the partial derivatives in Eq.~\eqref{eq:vibrational-mass}
are evaluated using the local generator $\hat{P}_i(s)$
of
the LQRPA solution
  as 
\begin{align}\label{eq:partial-derivative}
	\frac{\partial s_m}{\partial q_i} &=  \frac{\partial }{\partial q_i}
   \langle \phi(s) | \hat{s}_m|\phi(s)\rangle \notag \\
   &=  \langle \phi(s) | [\hat{s}_m, \frac{1}{i}\hat{P}_{i}(s)]|\phi(s)\rangle ,
\end{align}
%
  which 
is calculable
without numerical derivatives.

  \section{Finite-amplitude method for collective inertia}


In this work, we use the LQRPA method to
evaluate the collective inertia along a fission path.
Since it is computationally hard to solve the LQRPA equations for deformed nuclear shapes with the Skyrme EDF,
we employ the FAM to efficiently solve the LQRPA equations.
In this section, we explain our 
  method 
to obtain the expression of 
the collective inertia based on the FAM and LQRPA.

\subsection{Finite-amplitude method}

Here, the FAM
is 
  recapitulated.
The details of its derivation can be found in Refs. \cite{nakatsukasa07,avogadro11}.
The FAM equations can be expressed in the quasiparticle basis as
\begin{subequations}\label{eq:FAM}
\begin{align}
(E_\mu + E_\nu -\omega)X_{\mu\nu}(\omega) + \delta H^{20}_{\mu\nu}(\omega) &= -F^{20}_{\mu\nu} \; ,\\  
(E_\mu + E_\nu +\omega)Y_{\mu\nu}(\omega) + \delta H^{02}_{\mu\nu}(\omega) &= -F^{02}_{\mu\nu}\; ,
\end{align}\end{subequations}
where 
$X_{\mu\nu}(\omega)$ and $Y_{\mu\nu}(\omega)$ are the FAM amplitudes at a given frequency $\omega$, and
$\delta H^{20(02)}_{\mu\nu}$ and $F^{20(02)}_{\mu\nu}$ are two-quasiparticle components
of 
one-body induced 
field
$\delta \hat{H}$ and
an
external field $\hat{F}$, respectively.
%
%
The FAM equations \eqref{eq:FAM} are iteratively solved
at each $\omega$ until converged $X$ and $Y$ amplitudes are obtained.
Complex frequency $\omega$ is usually used with the imaginary part
corresponding to a smearing width.
Only one-body quantities (induced densities and induced
fields)
are necessary to solve the FAM equations.
%

The FAM strength function is given from converged $X$ and $Y$ amplitudes as 
\begin{equation}\label{eq:strengthXY}
  S(\hat{F},\omega) = \sum_{\mu<\nu}\left[F_{\mu\nu}^{20*}X_{\mu\nu}(\omega)+F_{\mu\nu}^{02*}Y_{\mu\nu}(\omega) \right].
\end{equation}
Taking the frequency $\omega$ real and positive, the transition strength distribution is given as
%
\begin{equation}\label{eq:transitionFAM}
  \frac{dB(\hat{F},\omega)}{d\omega}
	 \equiv \sum_{i(\Omega_i >0)} \left|\bra{i}\hat{F}\ket{0}\right|^2 \delta(\omega-\Omega_i)
	= -\frac{1}{\pi}\text{Im} S(\hat{F},\omega),
\end{equation}
where $\ket{0}$ is the QRPA vacuum and $\ket{i}\equiv\hat{O}_i^\dagger\ket{0}$ 
is a state of the QRPA normal mode of excitation
with the eigenfrequency $\Omega_i$.
%

The relation between the FAM strength function \eqref{eq:strengthXY}
and the QRPA transition strength
  $|\langle i|\hat{F}| 0\rangle|^2$ 
between the ground state and a QRPA eigenstate
$\ket{i}$
is given as \cite{hinohara13}
\begin{equation}
	S(\hat{F},\omega) = -\sum_{i (\Omega_i >0)}
	\left(\frac{|\langle i|\hat{F}|0\rangle|^2}{\Omega_i-\omega}
  +\frac{|\langle 0|\hat{F}|i\rangle|^2}{\Omega_i + \omega} \right) .
\end{equation}
%

\subsection{FAM in the PQ representation}

In Ref.~\cite{hinohara15b},
the FAM in the PQ representation of the QRPA 
was formulated to investigate the NG modes
and the Thouless--Valatin inertia of the NG modes.
In this subsection, we summarize the formulation of the FAM in the
PQ representation.


In the PQ representation of the QRPA, 
the
coordinate and the conjugate momentum
operators, $\hat{Q}_i$ and $\hat{P}_i$, 
  which are both Hermitian, 
are constructed
from the QRPA phonon operators
  $(\hat{O}_i,\hat{O}^\dagger_i)$ 
as
\begin{align}
  \hat{Q}_i &= \sqrt{\frac{1}{2M_i\Omega_i}}(\hat{O}_i + \hat{O}^{\dagger}_i),
	\label{eq:Q_i}\\
  \hat{P}_i &= \frac{1}{i}\sqrt{\frac{M_i\Omega_i}{2}}(\hat{O}_i - \hat{O}^{\dagger}_i),
	\label{eq:P_i}
\end{align}
where $M_i$
is 
the inertia parameter.
The operators $\hat{Q}_i$ and $\hat{P}_i$ fulfill the following
commutation relations, 
\begin{equation}
  \bra{0} [\hat{Q}_i,\hat{P}_j] \ket{0}= i \delta_{ij}, \;
  \bra{0} [\hat{Q}_i,\hat{Q}_j] \ket{0} = \bra{0} [\hat{P}_i,\hat{P}_j] \ket{0} = 0, 
\label{eq:weak_canonicity_condition}
\end{equation}
%
which guarantee the orthogonality among
different normal modes, however, their scale is still arbitrary as
$(\alpha\hat{Q}_i,\alpha^{-1}\hat{P}_i)$.
In the present study,
this scale is fixed by imposing the additional condition, $M_i=1$.

Using two-quasiparticle components $P_{\mu\nu}^{i}$ and $Q_{\mu\nu}^{i}$ of the
operators
$\hat{P}_i$ and $\hat{Q}_i$,
the FAM $X$ and $Y$ amplitudes
are given as
%
\begin{subequations}\label{eq:FAMamplitude-xy} 
\begin{align}
  X_{\mu\nu}(\omega) =& \sum_i \frac{1}{\omega^2 -\Omega^2_i}
  \Bigl[(-i\omega P_{\mu\nu}^i + \Omega_i^2 Q_{\mu\nu}^i)
	q_i(\hat{F}) \notag \\
        &\qquad + (P_{\mu\nu}^i + i\omega Q_{\mu\nu}^i)
	p_i(\hat{F}) \Bigr],\\
  Y_{\mu\nu}(\omega) =& \sum_i \frac{1}{\omega^2 -\Omega^2_i}
  \Bigl[(i\omega P_{\mu\nu}^{i*} - \Omega_i^2 Q_{\mu\nu}^{i*})
	q_i(\hat{F}) \notag \\
        &\qquad + (-P_{\mu\nu}^{i*} - i\omega Q_{\mu\nu}^{i*})
	p_i(\hat{F}) \Bigr],
\end{align} \end{subequations}
%
where
        $p_i(\hat{F})$ and $q_i(\hat{F})$
	are defined as
\begin{align}
	p_i(\hat{F})
	& \equiv \bra{0} [\hat{P}_i, \hat{F}]\ket{0}
  = \sum_{\mu<\nu} \left(P_{\mu\nu}^{i*} F_{\mu\nu}^{20}-P_{\mu\nu}^{i} F_{\mu\nu}^{02}\right), \label{eq:transitionP} \\
	q_i(\hat{F})
	& \equiv \bra{0} [\hat{Q}_i, \hat{F}]\ket{0}
  = \sum_{\mu<\nu} \left(Q_{\mu\nu}^{i*} F_{\mu\nu}^{20}-Q_{\mu\nu}^{i} F_{\mu\nu}^{02}\right). \label{eq:transitionQ}
\end{align}
%
%
%
The FAM strength function \eqref{eq:strengthXY}
is then
rewritten 
as
\begin{align}
  S(\hat{F},\omega) &= \sum_{\mu < \nu} \left[F_{\mu\nu}^{20*} X_{\mu\nu}(\omega)+ F_{\mu\nu}^{02*} Y_{\mu\nu}(\omega)\right]  \notag \\
  &= \sum_i \frac{1}{\omega^2 -\Omega^2_i}
  \left[
	|p_i(\hat{F})|^2
        + \Omega_i^2
	|q_i(\hat{F})|^2 + \omega \; r_i(\hat{F})
	\right],
  \label{eq:strengthPQ}
\end{align}
%
with a real quantity
\begin{equation}
	r_i(\hat{F}) 
	\equiv  i \left[
	q_i^*(\hat{F}) p_i(\hat{F}) - q_i(\hat{F}) p_i^*(\hat{F}) 
	\right].
\end{equation}
The FAM $X(\omega)$ and $Y(\omega)$ amplitudes~\eqref{eq:FAMamplitude-xy}
and the FAM strength function \eqref{eq:strengthPQ}  
are defined 
in the whole complex plane 
$\omega$ except for $\omega=\pm \Omega_i$.
  They are well defined even with 
the presence of
the NG modes ($\Omega_i^2=0$) and the imaginary solutions ($\Omega_i^2<0$)
which can appear in 
the
LQRPA 
at
nonequilibrium CHFB states.
%

\subsection{Collective inertia with FAM plus contour integration technique}

For simplicity, in this paper, we
assume 
only one collective coordinate $q$, and then
adopt 
the isoscalar quadrupole moment
$\hat{s} = \hat{Q}_{20}= \sqrt{16\pi/5}\sum_{i=1}^A r_i^2 Y_{20}(\hat{\boldsymbol{r}}_i)$
  as the constraint.
For the LQRPA calculation, we adopt 
the same operator for the external field $\hat{F}=\hat{Q}_{20}$.
Then,
by transforming the scale of the coordinate from $q$ to
$s=\bra{\phi(s)}\hat{Q}_{20}\ket{\phi(s)}$, 
the expression of the collective inertia~\eqref{eq:vibrational-mass}
becomes
\begin{equation}
	{\cal M} \equiv \frac{d q}{d s} \frac{d q}{d s} ,
\end{equation}
with 
	$dq/ds=(ds/dq)^{-1}$ and
%
\begin{align}
		\frac{d s}{d q}
	& =  \langle \phi(s) | [\hat{Q}_{20}, \frac{1}{i}\hat{P}_{i}]|\phi(s)\rangle \notag \\
  & = i
		p_i(\hat{Q}_{20}).
\end{align}
%
%
We will see below that $p_i(\hat{Q}_{20})$ is pure imaginary, leading to real $ds/dq$.
We select the normal mode $i$ which has the largest value of $|p_i(\hat{Q}_{20})|^2$
among the low-lying eigenmodes.

%
%
%
%
The remaining task is the calculation of $p_i(\hat{F})=p_i(\hat{Q}_{20})$
in the FAM.
From Eqs.~(\ref{eq:Q_i}) and (\ref{eq:P_i}),
$P^i$ and $Q^i$ are given in terms of the forward and backward amplitudes
of the QRPA normal modes, $X^i$ and $Y^i$, as
%
\begin{align}
  P_{\mu\nu}^i &= i \sqrt{\frac{\Omega_i}{2}} (X_{\mu\nu}^i + Y_{\mu\nu}^i), \\
  Q_{\mu\nu}^i &=   \sqrt{\frac{1}{2\Omega_i}}(X_{\mu\nu}^i - Y_{\mu\nu}^i).
\end{align} 
%
Note that we set $M_i=1$.
When the QRPA matrices are real, we may choose $X^i_{\mu\nu}$ and $Y^i_{\mu\nu}$ real.
This leads to real $Q^i_{\mu\nu}$ and pure imaginary $P^i_{\mu\nu}$.
%
In the present case, 
the external-field operator
  $\hat{F}=\hat{Q}_{20}$ 
is Hermitian and
their two-quasiparticle components are real,
namely, $F^{02}=F^{20*}=F^{20}$.
Thus,
  $q_i(\hat{F}) = 0$ and $p_i(\hat{F})=-p_i^*(\hat{F})$ (pure imaginary) 
hold according to Eq.~\eqref{eq:transitionQ}.
Then, the FAM strength function \eqref{eq:strengthPQ} becomes
\begin{equation}\label{eq:strengthP}
  S(\hat{F},\omega) = \sum_{i} \frac{1}{\omega^2 -\Omega^2_i}
	|p_i(\hat{F})|^2 .
\end{equation}
%
%
%
Since the right-hand side of Eq.~\eqref{eq:strengthP} has first-order poles at
$\omega = \pm \Omega_i$,
we obtain the following expression 
using 
Cauchy's integral formula,
\begin{equation}\label{eq:cont-wS}
  \frac{1}{2\pi i} \oint_{C_i} \omega S(\hat{F}, \omega) d\omega
  = \frac{1}{2}
	|p_i(\hat{F})|^2,
\end{equation}
where 
the 
contour circle $C_i$ in the complex energy plane
is chosen to encircle only
the 
pole 
$\omega=\Omega_i$.
The strength functions, $S(\hat{F}, \omega)$,
at different values of $\omega$ 
are obtained as Eq.~(\ref{eq:strengthXY})
with iterative solution of 
  Eq.~\eqref{eq:FAM}. 

%
The 
eigenfrequency $\Omega_i$
can be also obtained 
by
combining Eq.~(\ref{eq:cont-wS}) with 
the following contour integration,
\begin{equation} \label{eq:cont-S}
  \frac{1}{2\pi i} \oint_{C_i}  S(\hat{F}, \omega) d\omega
  = \frac{1}{2\Omega_i} 
  |p_i(\hat{F})|^2.
\end{equation}
%
The expressions \eqref{eq:cont-wS} and \eqref{eq:cont-S}
can be used for both
real and pure imaginary eigenfrequencies.

\subsection{Numerical procedure}
\label{sec:numerical-procedure}



To prepare the CHFB states along the fission path, 
we solve the CHFB equations with the two-basis method~\cite{gall94,terasaki95},
%
where during iteration the HFB Hamiltonian is diagonalized in a single-particle basis
that converges to the eigenstates of the mean-field Hamiltonian,
called the Hartree-Fock (HF) basis,
and the local densities are constructed
in the canonical basis that diagonalizes the density matrix.
The single-particle wave functions and fields are represented in
a three-dimensional (3D) Cartesian mesh.
Nuclear shapes are restricted to have three plane reflection symmetries
about the $x = 0$, $y = 0$, and $z = 0$ planes.
This reduces 
the model space
to 1/8 of the full box
and can be realized by choosing the single-particle wave functions as 
eigenstates of parity, $z$ signature, and $y$ time simplex
\cite{bonche85,bonche87,ev8,hellemans12}.
This symmetry restriction prevents us from treating asymmetric fission in
the present study.
%
We use a numerical box of 13.2\,fm for $x$ and $y$ directions
and 19.6\,fm for $z$ direction in $x>0$, $y>0$, and $z>0$ 
in order to express prolately deformed fissioning shapes,
discretized with the mesh size of $0.8$\,fm for all the directions.
The number of mesh points then becomes $16\times 16\times 24=6144$.
The single-particle basis consists of 2460 neutron and 1840 proton HF-basis states
to achieve the maximum quasiparticle energy of $E_{\text{QP}}^{\text{max}}\approx 60$\,MeV
for both neutrons and protons in $^{240}$Pu and $^{256}$Fm.
Constrained quantities are the isoscalar quadrupole moment $Q_{20}=\langle \hat{Q}_{20}\rangle$
with $\langle \hat{Q}_{22}\rangle =0$, keeping axially symmetric shapes.
The symmetry restriction automatically constrains the position of the center of mass, and
$\langle xy \rangle = \langle yz \rangle = \langle zx \rangle=0$.
The other even multipole moments are not constrained,
and their values are optimized to minimize the total energy,
while the expectation values of all odd multipole moments are kept to be zero
due to the plane reflection symmetries of nuclear shapes.
We used the SkM$^*$ EDF \cite{bartel82} and 
the contact volume pairing with a pairing window of 20\,MeV above and below
the Fermi energy in the HF basis described in Refs.~\cite{bonche85,ev8}
to avoid divergence in the pairing energy.
The pairing strengths are adjusted separately for neutrons and protons in order  
to reproduce the empirical pairing gaps in $^{240}$Pu or in $^{256}$Fm.

We have extended our 3D FAM-QRPA code developed in Ref.~\cite{washiyama17}
to perform FAM calculations 
  at 
CHFB states and
the FAM-QRPA with the contour integration.
We include in the FAM the full quasiparticle basis included in the CHFB calculations
to satisfy full selfconsistency between the CHFB and LQRPA calculations.
We employ the modified Broyden method~\cite{baran08}
for iteratively solving the FAM equations.
%

To obtain the 
quantity 
  $|p_i(\hat{F})|^2$ 
by the contour integration~\eqref{eq:cont-wS},
it is necessary to know in advance an approximate position of
each QRPA pole $\omega=\Omega_i$
in the complex energy plane
for determining an appropriate contour $C_i$.
The QRPA poles are expected to appear as sharp peaks in the FAM strength distribution.
Therefore, we first calculate the FAM strength distribution~\eqref{eq:transitionFAM}
in the low-lying states with $\omega^2 < 16$\,MeV$^2$.
%
For each pole,
we calculate
$|p_i(\hat{F})|^2$ 
and
adopt the most collective mode with the largest value of $|p_i(\hat{F})|^2$.
  A detail of our procedure of selecting the most collective mode
  in the FAM 
  is given in Appendix.
%
For 
the 
integration contour,
we use a circle centered at the estimated peak frequency.
The radius is 
0.05\,MeV for real solutions,
and 
0.1\,MeV for an imaginary solution.
If 
there are
other 
poles close to the present pole,
we use a circle with even smaller radius.
The number of discretization points along the circle
is 12 for real and 8 for imaginary solutions.

We perform numerical calculations of the FAM with a hybrid parallel scheme;
The MPI parallelization is adopted for calculation of $S(\hat{F},\omega)$ at
different $\omega$ points.
The FAM iterative solution of Eq.~\eqref{eq:FAM}
at each $\omega$ is performed using the OpenMP parallel calculation.
It takes about 160 
core hours on 
  Oakforest-PACS 
for the FAM contour integration for a real solution at one deformation point
with the present model space.


\begin{figure}
  \includegraphics[width=0.95\linewidth]{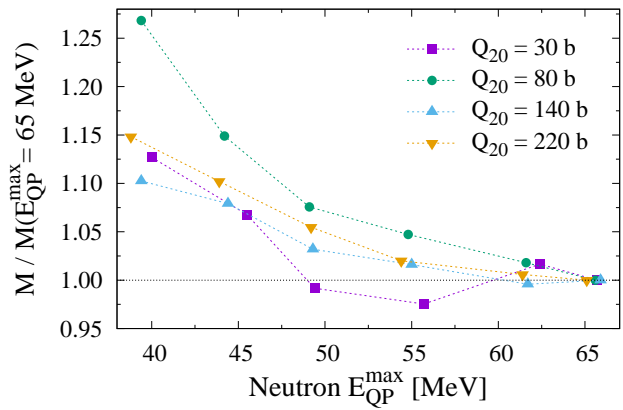}%
  \caption{\label{fig:convergence}
    Collective inertia as a function of $E_{\text{QP}}^{\text{max}}$ for neutrons 
    divided by that with neutron $E_{\text{QP}}^{\text{max}}\approx 65$\,MeV 
    for four different deformation points in ${}^{240}$Pu.
    Included neutron HF-basis states are 1200, 1440, 1632, 1998, 2460, and 2700
    for $E_{\text{QP}}^{\text{max}}\approx 40$, 45, 50, 55, 60, and 65\,MeV, respectively.
}\end{figure}
%

Before we proceed,
we would like to discuss the convergence property of the collective inertia
with respect to the size of the model space included in the ground state.
%
%
To check the convergence property, we calculated the collective inertia
with different numbers of HF-basis states corresponding to
different $E_{\text{QP}}^{\text{max}}$ values in the ground state.
Figure~\ref{fig:convergence} shows the collective inertia as a function of
neutron $E_{\text{QP}}^{\text{max}}$ (nearly equal to proton $E_{\text{QP}}^{\text{max}}$)
divided by that calculated with neutron $E_{\text{QP}}^{\text{max}} \approx 65$\,MeV
as a reference value for four 
various deformation points in ${}^{240}$Pu.
We find that
the model space with neutron
$E_{\text{QP}}^{\text{max}} \ge 55$\,MeV 
gives a good convergence within about less than 5\% to the value
with $E_{\text{QP}}^{\text{max}} \approx 65$\,MeV for the four various deformation points.
Therefore, 
as mentioned above, we include 2460 neutron and 1840 proton HF-basis states
corresponding to 
$E_{\text{QP}}^{\text{max}}\approx 60$\,MeV
in our calculations.
%
Note again that we do not introduce in the FAM calculations
the additional 
two-quasiparticle energy cutoff,
which has been usually employed to reduce the dimension of the QRPA matrix.


\section{Results}

Mass distributions of fission fragments measured in low-energy fission
show a sudden change 
from asymmetric to symmetric mass distributions when the neutron number changes.
A well known example is Fm isotopes,
where $^{256}$Fm shows an asymmetric mass distribution,
while $^{258}\text{Fm}$ shows that the main component
of fission fragments is a symmetric one~\cite{flynn72, hoffman80}.
Even more complicated multi-mode or multi-channel fission has been
observed and analyzed 
in the total-kinetic-energy distributions of fission fragments
in actinides~\cite{hulet86,brosa86}.
What determines the mass distribution?
It is still an open question, and the fragment mass distribution is
important for the fission recycling of the $r$-process as well.
Since, so far, most of studies are focused on the potential landscape,
we aim at investigating effect of the 
  collective inertia. 
In this section, we take $^{240}$Pu and $^{256}$Fm as examples
to show importance of dynamical residual effects on the inertia.
Although, in this paper, 
the calculation is performed only along the symmetric fission path,
we will show that the residual effect on the inertia may hinder
the symmetric fission probability.

\subsection{Collective inertia along a symmetric fission path in ${}^{240}$Pu}

\begin{figure}
  \includegraphics[width=0.95\linewidth]{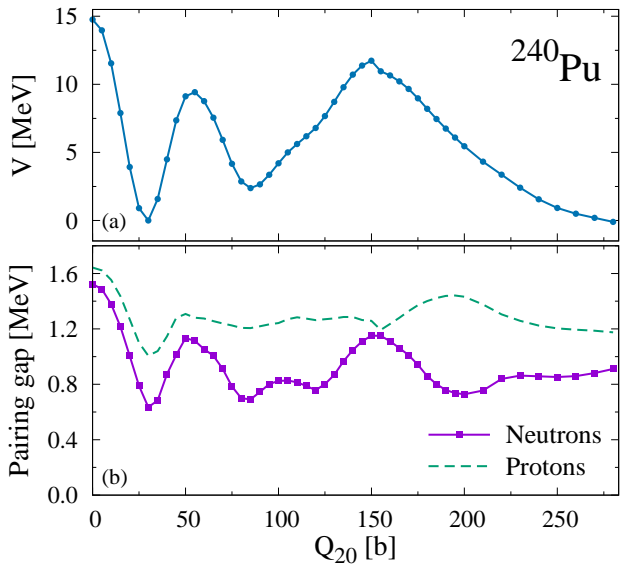}%
  \caption{\label{fig:pot.gap.Pu240}
    (a) Potential energy  
      (b) pairing gaps for
    neutrons (solid line) and protons (dashed line) as a function of
    quadrupole moment $Q_{20}$ in units of barn in ${}^{240}$Pu. }
\end{figure}

Figure~\ref{fig:pot.gap.Pu240}(a) shows the collective potential energy 
as a function of expectation value of quadrupole moment $Q_{20}$ in units of barn (b)
along a symmetric fission path 
in $^{240}\text{Pu}$ obtained from the CHFB calculation.
The ground state of ${}^{240}$Pu is found at $Q_{20} \approx 30$\,b.
The first fission barrier, 
the 
fission isomer (local minimum), and
the 
second fission barrier are obtained at $Q_{20} \approx 55$\,b, $Q_{20} \approx 85$\,b,
and $Q_{20} \approx 150$\,b, respectively.
We would like to note that
the heights of the first and second fission barriers would become lower
if we could include triaxial deformation for the first fission barrier 
and octupole deformation for the second fission barrier~\cite{bonneau04,zhou12}. 
Neutron and proton pairing gaps are shown along the fission path
in Fig.~\ref{fig:pot.gap.Pu240}(b).
The neutron pairing gap becomes
larger at the regions near the fission barriers and
smaller near the ground state and fission isomer.
The proton pairing gap shows a moderate behavior
because of the difference between neutron and proton single-particle structures.
It is known that transition strengths of low-lying modes
are affected by the strength of pairing correlations.

\begin{figure}
  \includegraphics[width=\linewidth]{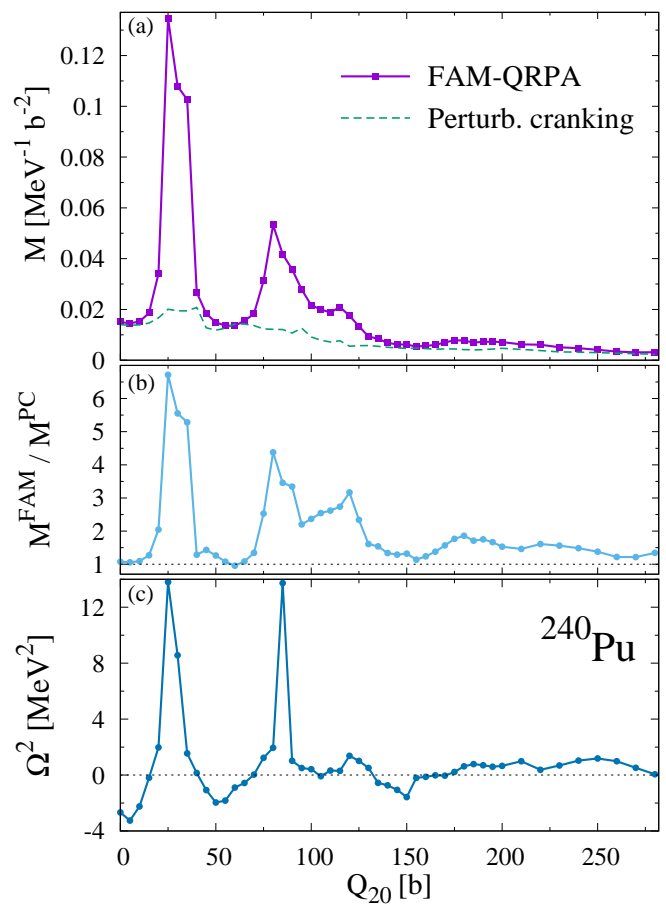}%
  \caption{\label{fig:mass.omega.Pu240}
    (a) Collective inertia of the FAM-QRPA calculation
    shown by the filled-square solid line,
    and of the perturbative cranking approximation by the dashed line,
    (b) ratio of the FAM-QRPA inertia to the perturbative cranking inertia,
    (c) squared QRPA eigenfrequency as a function of $Q_{20}$ in ${}^{240}$Pu.
  }
\end{figure}
%

Figure~\ref{fig:mass.omega.Pu240}(a) shows the collective inertia
obtained with the present FAM-QRPA calculation by the filled-square solid line.
This shows two prominent peaks at $Q_{20}\approx 25$\,b and $Q_{20} \approx 80$\,b
and a sudden change near these peaks.
These states closely correspond to the ground state and the fission isomer
observed in Fig.~\ref{fig:pot.gap.Pu240}(a)
and to the states where pairing becomes weak in Fig.~\ref{fig:pot.gap.Pu240}(b).
For comparison, we add in Fig.~\ref{fig:mass.omega.Pu240}(a)
the collective inertia obtained by the perturbative cranking approximation.
It is clearly shown that
the FAM-QRPA inertia is larger than the perturbative cranking inertia.
We emphasize this point in Fig.~\ref{fig:mass.omega.Pu240}(b),
showing that the ratio of the FAM-QRPA inertia to the perturbative cranking inertia
  always 
exceeds unity,
except at $Q_{20}=60$\,b.
This indicates that the action integral in the WKB approximation
will be significantly affected by the increase
of the collective inertia
near 
the fission barriers.
The behaviors of the two inertias are different;
the FAM-QRPA inertia shows a strong
variation, 
while the perturbative cranking inertia varies smoothly as $Q_{20}$ increases.

We estimate the action integral in the WKB approximation using
the obtained collective inertia and potential along the symmetric fission path.
The action integral reads
\begin{equation}
  S = \int_{Q_{\text{in}}}^{Q_{\text{out}}} dQ_{20} \sqrt{2\mathcal{M}(Q_{20})[V(Q_{20})-E_0]},
\end{equation}
where $Q_{\text{in}}$ and $Q_{\text{out}}$ are the classical inner and outer turning points,
and $E_0$ is the HFB ground state energy.
  We obtain the action integrals $S_{\text{FAM}}=82.0$ for the FAM-QRPA inertia
and $S_{\text{PC}} =62.0$ for the perturbative cranking inertia  
from $Q_{\text{in}} = 30$\,b (the ground state) to $Q_{\text{out}}=270$\,b.
This difference leads to many orders of magnitude difference in the fission half-life.

Figure~\ref{fig:mass.omega.Pu240}(c) shows
the eigenfrequency squared $\Omega^2$ 
of 
the LQRPA solution selected as 
the most collective mode at each deformation.
This represents the curvature of the collective potential,
which can indeed become negative at the fission barriers.
Larger values of $\Omega^2$ are found at the states
with larger values of the collective inertia.
%
As $\Omega^2$ becomes larger,
the ratio of the FAM-QRPA inertia to the perturbative cranking inertia
becomes larger. 
When $\Omega^2$ are negative, the ratio is close to unity.


\begin{figure}
  \includegraphics[width=0.95\linewidth]{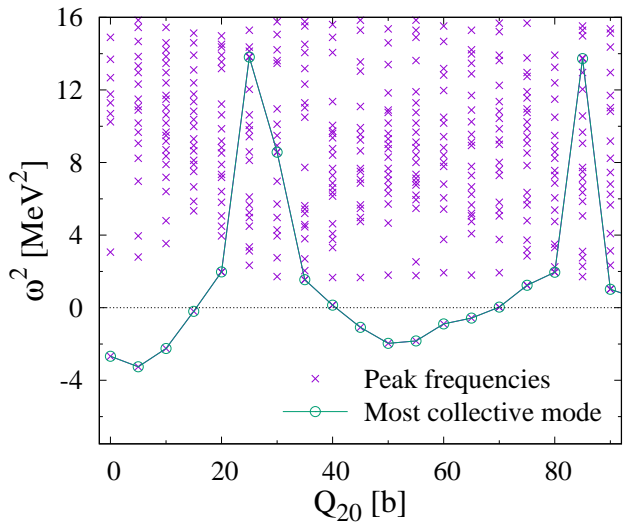}%
  \caption{\label{fig:peaks}
    Squared peak frequencies identified from the FAM strength distribution
    shown by the crosses as a function of $Q_{20}$ in $^{240}$Pu.
    Eigenfrequency squared of the most collective mode is
    shown by the open-circle solid line.
}\end{figure}

In Fig.~\ref{fig:peaks},
we plot the squared peak frequencies identified from the FAM strength distribution
in $\omega^2 < 16\,\text{MeV}^2$
as a function of $Q_{20}$ up to the fission isomer ($Q_{20} \le 90$\,b).
Note that, because of a finite smearing width ($=0.01$\,MeV) used in the FAM calculation,
the identified peak frequencies may slightly differ from
the QRPA eigenfrequencies and may not correspond to all the eigenfrequencies.
In the figure, 
the eigenfrequency squared of the most collective mode
that is used in the FAM-QRPA collective inertia
is shown by the open-circle solid line.
At most regions in $Q_{20}$,
we find that the most collective mode 
corresponds to the lowest-frequency mode
and 
  is well separated 
  from other peaks. 
On the other hand, around the ground state ($Q_{20}\approx 30$\,b)
and the fission isomer ($Q_{20}\approx 85$\,b), where pairing becomes weak,
  we find that the most collective mode appears at frequencies close to other peaks,
  or at higher frequencies. 
The increase of the FAM-QRPA inertia near the ground and the fission isomer states
is due to decrease of the quadrupole collectivity which leads to decrease in $|p_i(\hat{Q}_{20})|^2$.
In fact, near the ground state,
we find that the character of the lowest frequency mode changes from
the quadrupole vibration to the pair vibration.
%

It is known that QRPA eigenmodes that possess pair-vibrational character
tend to appear in low-frequency regions when pairing becomes weak.
%
%
  We 
analyze the character of the lowest-frequency mode
at $\omega \sim 1.5$\,MeV at $Q_{20}=25$, 30, and 35\,b
by using the FAM calculation with the pair-vibrational field
as an external field.
The 
lowest-frequency 
modes 
at $Q_{20}=25$, 30, and 35\,b
have 
a strong neutron pair-vibrational character
and approximately satisfy the relation $\Omega \approx 2\Delta_n$,
where $\Delta_n$ denotes the neutron pairing gap.
%
%
%
The 
  lowest-frequency modes at $Q_{20}=25$ and 30\,b
  are not the most collective
  in the quadrupole $Q_{20}$ nature.
  We also 
  confirm this 
  by calculating the collective inertia
  for 
  those lowest modes.
  For the lowest mode, the collective inertia would be
  $\mathcal{M}=0.190\,\text{MeV}^{-1}\text{b}^{-2}$ at $Q_{20}=25$\,b
  and
  $\mathcal{M}=1.02\,\text{MeV}^{-1}\text{b}^{-2}$ at $Q_{20}=30$\,b.
  These values are significantly larger than the values shown
  in Fig.~\ref{fig:mass.omega.Pu240}(a),
  indicating the character change of the lowest mode from
  the quadrupole vibration to the neutron pair vibration.


\subsection{Comparison between the FAM-QRPA and nonperturbative cranking inertias in ${}^{256}$Fm}

%

The aim of taking the ${}^{256}$Fm case
is to compare the FAM-QRPA collective inertia with that
of 
the nonperturbative cranking approximation by Baran et al. in Ref.~\cite{baran11}.
In ${}^{256}$Fm, it is known that
reflection-symmetric and reflection-asymmetric fission paths bifurcate
at $Q_{20} \approx 130$\,b \cite{warda02,dubray08,staszczak09}.
In Ref.~\cite{baran11},
the collective inertia was calculated along
a reflection-symmetric fission path in $Q_{20}<130$\,b
and along a reflection-asymmetric fission path in $Q_{20}>130$\,b.
As we do not include reflection-asymmetric fission paths,
we compare the FAM-QRPA collective inertia with that in Ref.~\cite{baran11} 
along the reflection-symmetric fission path in $Q_{20}\le 130$\,b.

\begin{figure}
  \includegraphics[width=0.95\linewidth]{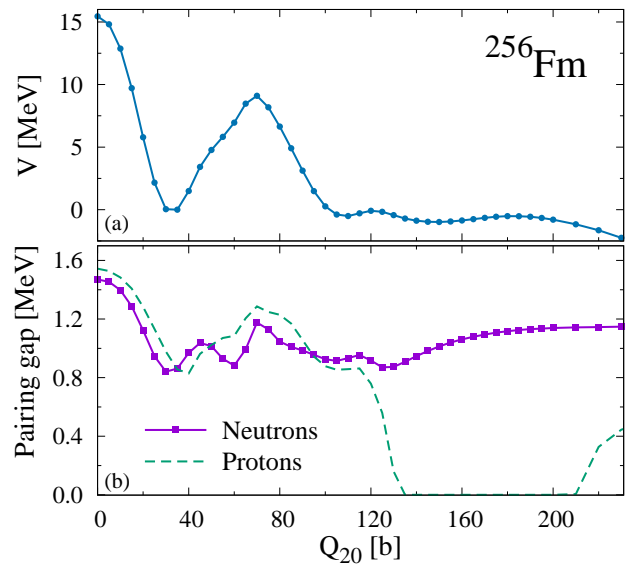}%
  \caption{\label{fig:pot.gap.Fm256}
    Same as Fig.~\ref{fig:pot.gap.Pu240}, but for ${}^{256}$Fm.
}\end{figure}

First, we show in Fig.~\ref{fig:pot.gap.Fm256} the potential and pairing gaps 
along an axial and reflection-symmetric fission path as a function of $Q_{20}$.
The structure of a fission isomer at $Q_{20} \approx 110$\,b almost vanishes
due to low second fission barrier height of about 0.4\,MeV.
The proton pairing vanishes in $Q_{20} \ge 130$\,b.
These properties are consistent with previous EDF studies \cite{warda02,dubray08,staszczak09,baran11}.
\begin{figure}
  \includegraphics[width=\linewidth]{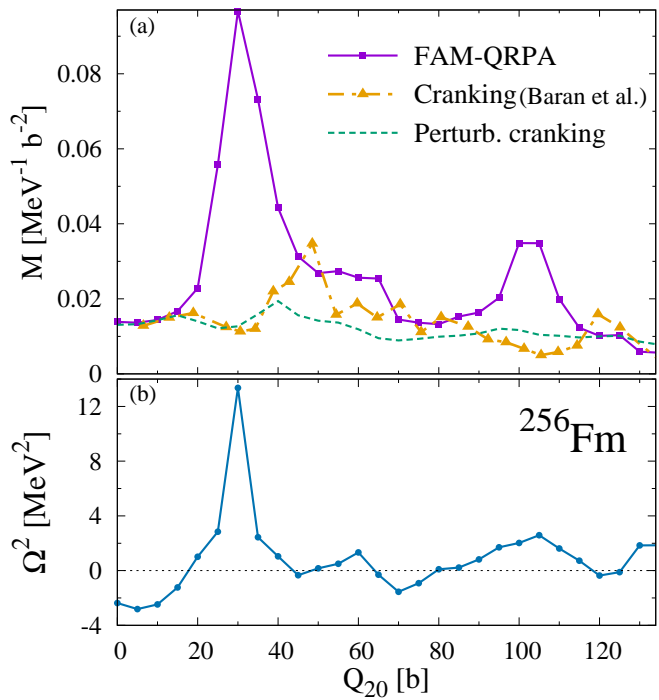}%
  \caption{\label{fig:mass.omega.Fm256}
    (a) Comparison between the FAM-QRPA inertia (the filled-square solid line),
    the nonperturbative cranking inertia taken from Baran et al.~\cite{baran11} (the filled-triangle dot-dashed line), 
    and the perturbative cranking inertia (dashed line), and
    (b) squared QRPA eigenfrequency as a function of $Q_{20}$
    along a symmetric fission path in ${}^{256}$Fm.
  }\end{figure}

Figure~\ref{fig:mass.omega.Fm256}(a) shows the FAM-QRPA collective inertia
as a function of $Q_{20}$.
Two peaks in the FAM-QRPA inertia are clearly seen
at the ground state ($Q_{20}\approx 30$\,b) and fission isomer ($Q_{20}\approx105$\,b),
which is similar to the case of ${}^{240}$Pu.
In this figure, we compare the FAM-QRPA inertia with
the inertia of the nonperturbative cranking approximation~\cite{baran11},
as well as the perturbative cranking one.
We
should 
note that
the model space, the pairing functional, and so on, adopted in the present study
are different from those in Ref.~\cite{baran11}. 
%
The FAM-QRPA inertia and the nonperturbative cranking inertia vary significantly
as $Q_{20}$ changes, compared with a smooth behavior of
the perturbative cranking inertia.
We find two significant differences between the FAM-QRPA inertia
and the nonperturbative cranking inertia.
One is the magnitude of the collective inertia;
the FAM-QRPA collective inertia is significantly larger than
the nonperturbative cranking one at most deformation points.
Similar values are obtained near the top of the fission barrier at $Q_{20}\approx 70$\,b
and near the bifurcation at $Q_{20}\approx 130$\,b.
The other difference is 
the position of peaks in $Q_{20}$
in the collective inertia.
The FAM-QRPA inertia peaks at slightly smaller $Q_{20}$
  than the cranking one does.
These significant differences are, at least partially, due to the residual time-odd fields
neglected in the cranking approximation.

Figure~\ref{fig:mass.omega.Fm256}(b) shows the squared QRPA eigenfrequency
obtained by the FAM-QRPA.
The peak structure of the squared QRPA eigenfrequency
is similar to that of the collective inertia,
which is also seen in the ${}^{240}$Pu case.
This suggests that a structure change of the lowest frequency mode
affects the collective inertia in $^{256}$Fm,
analogous to the case of $^{240}$Pu.

\section{Conclusion}

We have developed a
  feasible method of calculating 
the collective inertia based on the CHFB + LQRPA method with the Skyrme EDF.
This method includes time-odd components of dynamical residual effects to the collective inertia
in a selfconsistent way.
We efficiently calculated the QRPA transition strengths
by using the FAM and contour integration technique.
We applied this
method 
of evaluating the collective inertia along a symmetric fission path of $^{240}$Pu and $^{256}$Fm.
The results show that the dynamical residual effects significantly affect
the collective inertia and result in an enhancement of the collective inertia
compared with the perturbative cranking one.
The enhancement depends strongly on the deformation of the states.
This is a consequence of microscopic dynamical effects.
In the case of $^{256}$Fm,
we also compare the FAM-QRPA inertia with
the nonperturbative cranking collective inertia in Ref.~\cite{baran11}.
Both collective inertias show peak structures,
which are not seen
in the perturbative cranking inertia.
The values of the deformation at the peaks
in the inertia are different.
The FAM-QRPA inertia takes much larger values than nonperturbative one
around the potential minima.

For future works,
it is desirable to lift the symmetry restriction and to study
collective inertia along the asymmetric fission paths.
We also 
plan to extend the present
method 
to include two or more collective variables
for constructing collective inertia tensors
such as a simultaneous treatment of quadrupole and octupole moments
along multiple fission paths.
Systematic study of the dynamical residual effects on collective inertia
in fission in actinide and transactinide nuclei will be important
not only for deeper understanding of fission
but also for reliable evaluation of fission half-lives in $r$-process nucleosynthesis.
The present 
method 
can be also used
to describe large-amplitude collective dynamics
with the collective Hamiltonian method.
It would be interesting to compare the FAM-QRPA inertia
with the full ATDHFB inertia \cite{wen16}.

\begin{acknowledgments}
  The work was supported in part by QR Program of Kyushu University,
  by JSPS KAKENHI Grant Numbers JP18H01209, JP19H05142 and JP20K03964,
  and by JSPS-NSFC Bilateral Program for Joint Research Project
  on Nuclear mass and life for unravelling mysteries of $r$-process.
  Numerical calculations were performed in part using the COMA (PACS-IX)
  and Oakforest-PACS Systems through
  the Multidisciplinary Cooperative Research Program of the
  Center for Computational Sciences, University of Tsukuba.
\end{acknowledgments}

\appendix*
\section{Selection of the most collective mode}

%
\begin{figure}
  \includegraphics[width=\linewidth]{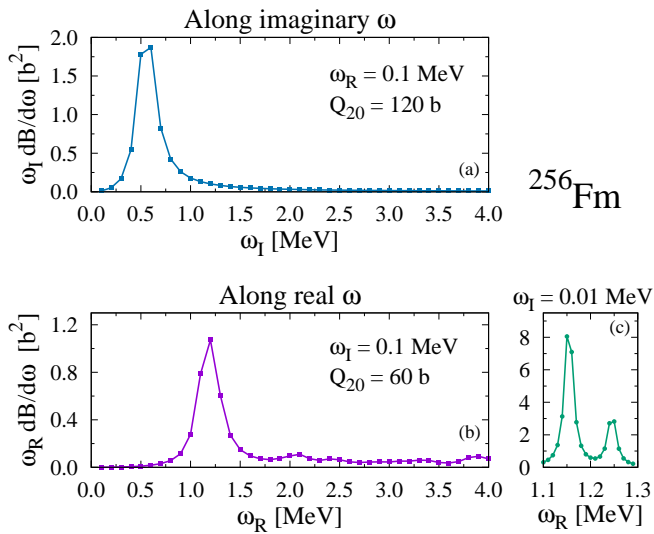}
  \caption{\label{fig:FAMstrength}
    Example of the strength distribution along the imaginary axis at $Q_{20}=120$\,b (a)
    and real axis at $Q_{20}=60$\,b (b) in $^{256}$Fm. These strength distributions are
    calculated at each 0.1\,MeV with a smearing width of 0.1\,MeV.
    (c) A smaller smearing width of 0.01\,MeV at each 0.01\,MeV in $\omega_R$
    is used for the same case as (b).
}\end{figure}

We explain our procedure of selecting the most collective QRPA eigenmode
to calculate the collective inertia. 
First, we perform the FAM calculation
with an isoscalar quadrupole external field $\hat{Q}_{20}$ and 
a complex frequency $\omega=\omega_R + i\omega_I$
along the imaginary and real $\omega$ axes
to obtain the FAM strength distribution.
For the FAM calculation along the imaginary $\omega$ axis,
$\omega_R$ has a role of a smearing width.

Next, we search sharp peaks in the FAM strength distribution,
which appear at approximate positions of the QRPA poles, $\omega = \Omega_i$,
in $0 \le \omega_R (\omega_I) \le 4$\,MeV along the real (imaginary) axis
with a smearing width of $\omega_I(\omega_R)=0.1$\,MeV.
An example of peak search from the strength distribution
is shown in Fig.~\ref{fig:FAMstrength} 
at $Q_{20}=60$ and 120\,b in $^{256}$Fm.
In Fig.~\ref{fig:FAMstrength}(a),
the strength distribution at $Q_{20}=120$\,b along the imaginary $\omega$ axis shows 
a clear peak with
large 
magnitude, corresponding to a pure imaginary QRPA pole.
We found that either single peak with
large 
magnitude or no peak appears
in the strength distribution along the imaginary $\omega$ axis
for the cases considered here.
In Fig.~\ref{fig:FAMstrength}(b),
for the strength distribution at $Q_{20}=60$\,b along the real $\omega$ axis,
one peak with larger magnitude is found at $\omega_R\approx 1.2$\,MeV.
However, we found that
this peak is resolved as two peaks at $\omega_R=1.15$ and $1.25$\,MeV
by the FAM calculation
with a smaller smearing width of $\omega_I=0.01$\,MeV,
which is illustrated in Fig.~\ref{fig:FAMstrength}(c).
To search sharp peaks in the strength distribution along the real $\omega$ axis,
we finally used $\omega_I=0.01$\,MeV to determine approximate positions of
the QRPA poles.
Note that with $\omega_I=0.01$\,MeV we can not resolve multiple peaks
within about 0.05 MeV in frequency in the strength distribution.

We 
take a few peaks from the
large 
peaks identified
in the FAM strength distribution along both the imaginary and real $\omega$ axes
as candidates of the most collective eigenmode.
Then, we perform the contour integration~\eqref{eq:cont-wS}
to obtain the transition strength for each candidate,
and
finally adopt 
the one with the largest
value of $|p_i(\hat{Q}_{20})|^2$ 
as the most collective eigenmode.

\end{document}